\documentclass{article}

\usepackage[english]{babel}

\usepackage[letterpaper,top=2cm,bottom=2cm,left=3cm,right=3cm,marginparwidth=1.75cm]{geometry}

\usepackage{amsmath,amssymb,amsfonts}
\usepackage{graphicx}
\usepackage{tabularx}
\usepackage{booktabs}
\usepackage{lscape}
\usepackage[colorlinks=true, allcolors=blue]{hyperref}

\title{Anomalous diffusion for mass transport phenomena II: Subdiffusion in polydimethylsiloxane (PDMS)}
\author{Nathaniel G. Hermann,\textsuperscript{1} Dmitry A. Markov,\textsuperscript{2} M. Shane Hutson\textsuperscript{1}* \\
\\
\textsuperscript{1}\textit{Department of Physics, Vanderbilt University, Nashville, TN, USA} \\
\textsuperscript{2}\textit{Department of Biomedical Engineering, Vanderbilt University, Nashville, TN, USA} \\
\textsuperscript{*}\textit{Corresponding author: shane.hutson@vanderbilt.edu}
}
\date{}

\begin{document}

\maketitle

\begin{abstract}
     Polydimethylsiloxane (PDMS) is a glassy polymer widely used in biomedical engineering, namely in microfluidics applications. However, PDMS is known to interact with hydrophobic chemicals. This interaction is exacerbated at the scale of microfluidics, making careful modeling of in-device concentrations vital for PDMS-based microfluidic devices. While it has been previously reported that many chemicals diffuse through PDMS, here we report that diffusion in PDMS is anomalous, \textit{i.e.} characterized by nonlinear, subdiffusive mean-squared displacements (MSD). We show that this anomalous diffusion can be modeled in the framework of stretched-time fractional diffusion, and report the transport parameters for a set of fluorescent tracer dyes. Depending on the device geometry and protocol, this anomalous behavior may have a significant impact, specifically in regards to cross-talk between microfluidic channels.
\end{abstract}

\section*{Introduction}
Polydimethylsiloxane (PDMS) is a viscoelastic polymer which is commonly used in biomedical devices. PDMS has many properties which are ideal for biomedical applications, including its low cost, optical transparency, flexibility, and biological compatibility \cite{Miranda2022}. However, PDMS is hydrophobic, and is known to interact with many hydrophobic chemicals \cite{Mcdonald2000,Toepke2006,Wang2012,VanMeer2017,Auner2019,Carlsen2020,Kemas2024,Hermann2025}. Many devices made using PDMS, especially microfluidic devices like organs-on-a-chip in which high surface-area-to-volume ratios make chemical interaction especially problematic. Therefore, one must either use mitigation strategies (either changing the composition of solution or the surface of PDMS) \cite{Fischer2019,Neves2024}, or be mass transport modeling to determine real in-device concentrations as a function of nominal inlet concentration. \\
\indent Prior studies have developed a mass transport model to describe chemical-PDMS interaction in terms of both partitioning into and diffusion within PDMS \cite{Hermann2025}. This models can, at relevant timescales, accurately replicate the concentration of a chemical within a PDMS microchannel. However, this model assumes that diffusion within PDMS is Fickian (\textit{i.e.} that it behaves according to Fick's second law). Anomalous, non-Fickian diffusion can occur in disordered media, including polymers, making the exclusion of this behavior potentially problematic for PDMS-based systems \cite{Muller-Plathe1992,Cuetos2018,Singh2020}. \\
\indent Anomalous diffusion is often characterized by its power-law-dependent mean-squared displacement (MSD). For a concentration profile $c(\mathbf{r},t)$,
\begin{equation}
    \text{MSD} = \int_{-\infty}^{\infty}\mathbf{r}\cdot\mathbf{r}\ c(\mathbf{r},t)d\mathbf{r}
\end{equation}
In a system with Fickian diffusion, one finds a linear relationship, $\text{MSD}\sim t$, while in one with anomalous diffusion, this relationship is characterized by an exponent $H$ such that $\text{MSD}\sim t^{H}$ \cite{Metzler2014}. This exponent is the Hurst exponent, which defines self-similarity in a system \cite{Gneiting2004}. This relationship links the observed ensemble behavior of MSD-characterized anomalous diffusion with the stochastic process driving single-particle behavior. Indeed, given a system such as the diffusion equation which describes the evolution of a probability density function (PDF) $f(\mathbf{r},t)$, one can define the equivalent stochastic process with identical marginal density function $f(\mathbf{r},t)$ \cite{Mainardi2010}. \\
\indent Given this one-to-one relationship, we will consider the ensemble behavior of small molecules diffusing through PDMS, and from this, determine the nature of the stochastic behavior of an individual molecule. To most fully generalize the system, we use the framework of stretched-time fractional diffusion (STF) to most generalize the system (as described in Part I). This ensemble description corresponds to the stochastic process of generalized gray Brownian motion (ggBm) \cite{Mura2009}. Characterizing anomalous diffusion in PDMS via STF diffusion is more than academic; it provides needed corrections for mass transport modeling of PDMS-based devices, particularly those at risk for cross-talk between closely-spaced microfluidic channels.

\section*{Methods}

\subsection*{PDMS preperation}
PDMS Sylgard 184 (Dow Corning, Auburn, MI) was mixed in a  10:1 mass ratio of elastomer base to curing agent. PDMS microchannels, measuring 21.1 mm in length, 1.5 mm in width, and 100 $\mu$m in height, were made by casting 6 mm of liquid PDMS over a SU-8 photoresist mold. The cast PDMS was then allowed to cure overnight in a  67°-C oven. After curing, PDMS was annealed for 4 hours in a 200°-C oven to stabilize mechanical properties \cite{Schneider2008}.

\subsection*{Dye solution preparation}
Fluorescent dyes spanning a wide range of physical parameters (\textit{e.g.} mass, volume, polarity, and hydrophobicity, Table \ref{DyeProperties}) were acquired as powders from Sigma Aldrich (St. Louis, MO). Stock solutions were prepared with 1x pH-7.4 phosphate buffered saline (PBS) (Thermo Fisher, Waltham, MA) to near maximum solubility. For chemicals with very low solubility in water, dimethyl sulfoxide (DMSO) was added to increase solubility. DMSO does not partition into nor interact with PDMS, making it a compatible organic solvent \cite{Lee2003}. 

\begin{table}
\centering
\renewcommand{\tabcolsep}{0.25cm}
\begin{tabular}{l c c c c c}
Dye Name & Mass (amu) & $\text{Vx}$ (mL/mol) & TPSA (\AA\textsuperscript{2}) & HBD & XLogP \\
\hline
\hline
Acridine Orange & $265.16$ & $2.18$ & $19.37$ & 0 & 2.03 \\
Coumarin 334 & $283.12$ & $2.09$ & $46.61$ & 0 & 1.58 \\
Coumarin 343 & $285.10$ & $2.01$ & $66.84$ & 1 & 1.74\\
Auramine O & $303.15$ & $2.52$ & $30.33$ & 1 & 2.99 \\
Coumarin 153 & $309.10$ & $1.98$ & $29.54$ & 0 & 2.66 \\
Coumarin 30 & $347.16$ & $2.66$ & $47.36$ & 0 & 3.11 \\
Crystal Violet & $407.21$ & $3.41$ & $9.49$ & 0 & 4.49 \\ 
Rhodamine B & $478.20$ & $3.75$ & $52.78$ & 1 & 4.74
\end{tabular}
\caption{\label{DyeProperties}Selected properties of fluorescent dyes used here, including molar mass, McGowan Volume ($\text{Vx}$), topological polar surface area (TPSA), H-bond donor count (HBD), and calculated logP (XLogP).}
\end{table}

\subsection*{Direct optical measurement of diffusion in PDMS}
Diffusion in PDMS was assayed via direct optical visualization. To do so, a solution of each fluorescent dye was loaded into a 21.1-mm long by 1.5-mm wide by 100-µm tall PDMS microchannel  and imaged for one hour using a 1× objective on a Nikon Ti2 Eclipse with X-light V2 spinning disk confocal microscope (Nikon Instruments, Melville, NY). After one hour, the microchannel was emptied and dried. The walls of the dry channel were then imaged for an additional 12 hours to follow the diffusive spread of any dye that had previously partitioned into the PDMS.

\subsection*{Molecular descriptor analysis}
Molecular descriptors were calculated for each dye using RDKit \cite{RDKit} (Listing S1). LASSO regression was then performed with $\log{D_{P}^{(\alpha\beta)}}$, $\alpha$, and $\beta$ as targets to determine which descriptors are correlated with STF diffusion parameters (Listing S2). LASSO regression was originally designed for linear regression, but we use it here primarily for parameter selection; here, it selects a subset of descriptors by forcing certain descriptor coefficients to zero \cite{Tibshirani2011}. Values of R\textsuperscript{2} are calculated for both the LASSO regression model and individual, univariate linear regression models  with descriptors having non-zero LASSO coefficients. A cross-validated R\textsuperscript{2} value is also calcualted for the LASSO regression model.

\subsection*{Modeling anomalous diffusion}
Concentration profiles of fluorescent dyes in PDMS first had numerical values of MSD calculated and fit in Wolfram Mathematica. These profiles were then directly fit to analytic solutions to fractional diffusion equations developed through the methods described in Part I of this work. Numerical approximations of the fractional complementary error function were constructed in Mathematica using the series form truncated at $n=100$ terms. 

\section*{Results and Discussion}
To characterize diffusion in PDMS, we used fluorescent intensity profiles  from a system of previously partitioned dye in PDMS imaged over 12 hours. To initially test for the occurrence of anomalous diffusion, we consider the MSD of these intensity (concentration) profiles. While this method cannot fully determine the parameters of STF diffusion, it can demonstrate the presence of non-linear dynamics.

\subsection*{Characterizing anomalous diffusion via MSD}
\begin{figure}
    \centering
    \includegraphics[scale = 1.25]{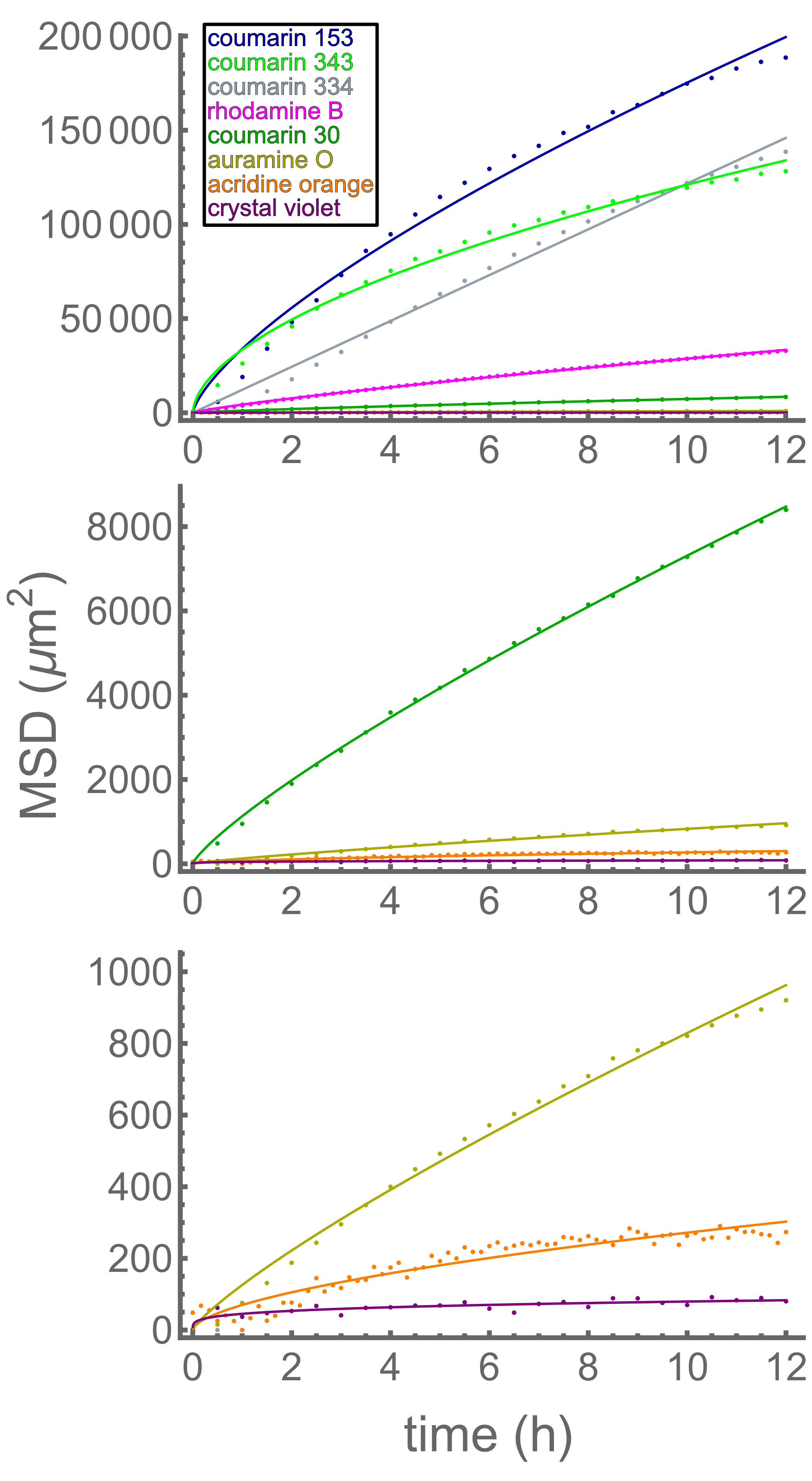}
    \caption{Plots of spreading MSD over time for eight fluorecent dyes.}
    \label{fig:MSDPlots}
\end{figure}

With the channel emptied, we treat our system of previously partitoned dye in PDMS as an effectively 1D system with a reflecting boundary at the empty channel wall. We imaged the concentration profile of each dye over 12 hours, and at each time $t$, we numerically calculate the MSD of each profile
\begin{equation}
    \text{MSD} = \frac{1}{2}\int_{0}^{\infty}x^{2}c(x,t)dx
\end{equation}
If diffusion is Fickian, we expect Gaussian concentration profiles,
\begin{equation}
    c(x,t) = \frac{1}{\sqrt{4\pi D_{P}t}}\exp{\left(-\frac{x^{2}}{4D_{P}(t+t_{0})}\right)}
\end{equation} 
while for STF diffusion, we expect concentration profiles given by
\begin{equation}
    c(x,t) = \frac{1}{\sqrt{4D_{P}^{(\alpha\beta)}t^{\alpha}}}M_{\beta/2}\left(\frac{|x|}{\sqrt{D_{P}^{(\alpha\beta)}}t^{\alpha}}\right)
\end{equation}
Here, $M$ is the M-Wright, or Mainardi, function, while $\alpha$ and $\beta$ are two characteristic exponents \cite{Mainardi2010}. The first, $\alpha$ is related to the self-similarity of this function, such that the Hurst exponent is $H=\alpha/2$. The second, $\beta$, defines the shape of the profile. Here, small values of $\alpha<1$ describe anomalous subdiffusion through non-linear temporal scaling, suggesting a physical slowing of molecules over time, which can arise for a multitude of reasons (\textit{e.g.} crowding, trapping, etc). Small values of $\beta<1$ describe anomalous subdiffusion through modifying the shape of the marginal density function of the stochastic process, suggesting heterogeneous media with local effects on particle motion. \\
\indent While the MSD of (3) is linear with respect to time, the MSD of Equation 4 is given by
\begin{equation}
    \text{MSD} = \frac{2}{\Gamma(\beta+1)}D_{P}^{(\alpha\beta)}t^{\alpha}
\end{equation}
By fitting the MSD to (5), we cannot indepdently determine the shape parameter $\beta$ or diffusion constant $D_{P}^{(\alpha\beta)}$,but we can detect anomalous diffusion by estimating $\alpha$. For the eight dyes used here, we find a wide variety of $\alpha$ values, from near-Fickian diffusion, $\alpha=0.998$ (coumarin 334) to extreme subdiffusion, $\alpha=0.245$ (crystal violet). These are listed in Table \ref{DyeParams}. Note that $\alpha$ describes the slowing of diffusion with time, not the overall spread of diffusion. For instance, coumarin 343 is much more subdiffusive ($\alpha=0.556$) than coumarin 334, but both reach MSD values $\approx 1.5\times 10^{6}$ \textmu m\textsuperscript{2} over 12 hours. Similarly, although auramine O and coumarin 30 have similar $\alpha$ values (0.820 amd 0.811 respectively), the MSD of the latter is 100x larger (10\textsuperscript{3} \textmu m \textsuperscript{2} versus 10\textsuperscript{3} \textmu m\textsuperscript{2}, Fig. \ref{fig:MSDPlots}).

\begin{table}
\centering
\renewcommand{\tabcolsep}{0.3cm}
\begin{tabular}{ l  c }
Dye Name & $\alpha$ \\
\hline
\hline
Crystal Violet & $0.245\pm0.047$ \\ 
Coumarin 343 & $0.556\pm0.018$ \\
Acridine Orange & $0.589\pm 0.034$ \\
Coumarin 153 & $0.712\pm 0.028$ \\
Coumarin 30 & $0.811\pm0.009$ \\
Rhodamine B & $0.820 \pm 0.004$ \\
Auramine O & $0.820\pm 0.025$ \\
Coumarin 334 & $0.998\pm 0.032$ \\
\end{tabular}
\caption{\label{DyeParams}Best fit value of anomalous diffusion order $\alpha$ for fluorescent dyes based on MSD.}
\end{table}

\subsection*{Characterizing anomalous diffusion through STF diffusion}
\begin{figure}
    \centering
    \includegraphics[scale = 0.95]{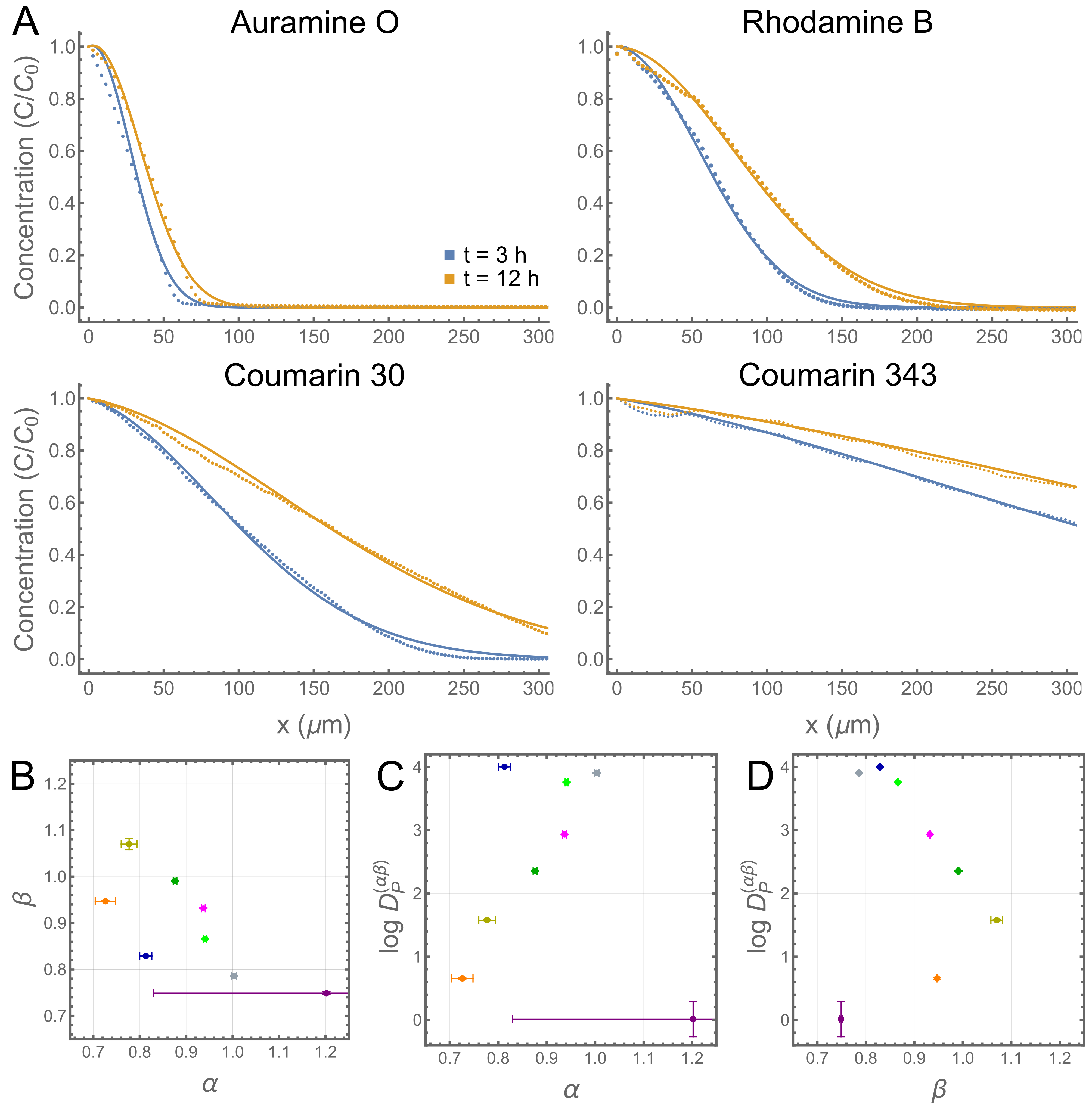}
    \caption{(A) Fits to STF diffusion solution for (clockwise from top-left) auramine O, rhodamine B, coumarin 343, and coumarin 30 at 3 and 12 hours after the start of imaging. (B-D) Parameter space of $\alpha$, $\beta$, and $\log{D_{P}^{(\alpha\beta)}}$.}
    \label{fig:STFFits}
\end{figure}

Given that 7 of these 8 dyes have clearly anomalous diffusion, we next fit the concentration profiles of previously partitioned dye PDMS directly to the analytic solution predicted for STF diffusion. Such anomalous diffusion is described by the master PDE:
\begin{equation}
    \frac{\partial c}{\partial t} = \frac{\alpha}{\beta}t^{\frac{\alpha}{\beta}-1}D_{P}^{\alpha\beta} \ _{RL}\mathcal{D}_{t^{\alpha/\beta}}^{1-\beta}\frac{\partial^{2}c}{\partial x^{2}}
\end{equation}
Here, $^{RL}\mathcal{D}$ is a Riemann-Liouville fractional derivative. For a delta function initial condition $c(x,0) = \delta(x)$, this PDE has solution
\begin{equation}
    c(x,t) = \frac{1}{\sqrt{4D_{P}^{(\alpha\beta)}t^{\alpha}}}M_{\beta/2}\left(\frac{|x|}{\sqrt{D_{P}^{(\alpha\beta)}t^{\alpha}}}\right)
\end{equation}
This system is discussed in Part I of this work. Given the difficulty of evaluating the M-Wright function directly at large arguments, we fit the concentration profiles of each dye to this form in three steps. First, we drop all times earlier than 3 hours and normalize at each time $t$ to the value $c(0,t)$ such that
\begin{equation*}
    \frac{c(x,t)}{c(0,t)} = M_{\beta/2}\left(\frac{|x|}{\sqrt{D_{P}^{(\alpha\beta)}(t+t_{0})^{\alpha}}}\right)
\end{equation*}
Dropping early times esures that the concentration profile is approximately M-Wright-like. This is a consequence of the Generalized Central Limit Theorem, which guarantees that for well-behaved but non-M-Wright initial conditions, a profile will nonetheless evolve into an M-Wright function over time (Appendix A). Second, at each time $t$, we fit $c(x,t)/c(0,t)$ to 
\begin{equation}
    \frac{c(x,t)}{c(0,t)} = M_{\beta/2}\left(\frac{|x|}{\sqrt{\sigma_{\alpha}(t)}}\right)
\end{equation}
For each dye, the best fit value of $\beta$ over time is tightly constrained, indicating that the process represented by $\beta$ is in fact uniform over time. We thus report the average value of$\beta$ over all time slices. Third and finally, we fit the values of $\sigma_{\alpha}(t)$ generated in the previous step to the function
\begin{equation}
    \sigma_{\alpha}(t) = \sqrt{D_{P}^{(\alpha\beta)}(t+t_{0})^{\alpha}}
\end{equation}
\indent We find that each of these dyes shows subdiffusive behavior with diffusivity spanning orders of magnitude (from 1 to 10\textsuperscript{4} \textmu m\textsuperscript{2}/t\textsuperscript{$\alpha$}). These fits (Fig. \ref{fig:STFFits}A) match the data well. None of the dyes tested herein show properly Fickian diffusion, and in fact many are well-fit to values of $\alpha\neq\beta$, indicating that the underlying stochastic process governing single particle motion is ggBm (see Part I). The best-fit values of $\alpha$, $\beta$, and $D_{P}^{(\alpha\beta)}$ are shown in Table \ref{STFFits}.

\begin{table}
\centering
\renewcommand{\tabcolsep}{0.3cm}
\begin{tabular}{l c c c}
Dye Name & $\log{D_{P}^{(\alpha\beta)}}$ (\textmu m\textsuperscript{2}/h\textsuperscript{$\alpha$}) & $\alpha$ & $\beta$ \\
\hline
\hline
Crystal Violet & $0.014\pm0.280$ & $1.202\pm0.372$ & $0.749\pm0.003$  \\ 
Acridine Orange  & $0.657\pm0.015$ & $0.726\pm0.022$ & $0.947\pm0.001$  \\
Auramine O & $1.578\pm0.012$ & $0.777\pm0.017$ & $1.070\pm0.012$ \\
Coumarin 30 & $2.355\pm0.003$ & $0.876\pm0.004$ & $0.991\pm0.001$ \\
Rhodamine B & $2.934\pm0.003$ & $0.937\pm0.004$ & $0.932\pm0.001$ \\
Coumarin 343 & $3.759\pm0.002$ & $0.941\pm0.003$ & $0.866\pm0.001$ \\
Coumarin 334  & $3.907\pm0.003$ & $1.003\pm0.005$ & $0.768\pm0.001$ \\
Coumarin 153 & $4.022\pm0.009$ & $0.813\pm0.013$ & $0.829\pm0.001$ \\
\end{tabular}
\caption{\label{STFFits}Fitted values to STF diffusion equation for fluorescent dyes.}
\end{table}

\subsection*{Exploratory study of molecular descriptor correlations with STF diffusion parameters}
To examine which molecular properties may correlate with anomalous diffusion in PDMS, we performed an exploratory study using RDKit to calculate molecular descriptors. For each dye, 208 descriptors were calculated. LASSO regression was then applied with STF parameters $\log{D_{P}^{(\alpha\beta)}}$, $\alpha$, and $\beta$ as targets to determine a subset of descriptors linearly-correlated to the target (Tables S1-3). For $\log{D_{P}^{(\alpha\beta)}}$, 7 descriptors were selected (MaxEStateIndex, MaxAbsEStateIndex, FpDensityMorgan1, SMR\_VSA5, SMR\_VSA6, VSA\_EState9, and fr\_NH0); for $\alpha$, 2 descriptors were selected (PEOE\_VSA7, VSA\_EState7); and for $\beta$, 8 descriptors were selected (MinAbsEStateIndex, BCUT2D\_LogPHI, PEOE\_VSA11, PEOE\_VSA12, EState\_VSA7, VSA\_EState7, NHOHCount, and fr\_NH2). The physical meanings of these descriptors are listed in Table S4. Univariate linear regression shows somewhat linear fits for $\log{D_{P}^{(\alpha\beta)}}$ (Fig. S1), while those for $\alpha$ and $\beta$ are of poorer quality (Fig. S2-3). For $\log{D_{P}^{(\alpha\beta)}}$, both LASSO regression R\textsuperscript{2} = 0.999 and cross-validated R\textsuperscript{2} = 0.725 indicate a fairly well-fit model, despite the extremely sparse data. On the other hand, $\alpha$ is fit to LASSO regression R\textsuperscript{2} = 0.465 and cross-validated R\textsuperscript{2} = -0.325 while $\beta$ is fit to LASSO regression R\textsuperscript{2} = 1 and cross-validated R\textsuperscript{2} = -0.343. These indicate non-predictive results, unsurprising given the small size of the dataset. 
Nevertheless, for the purposes of an exploratory study, we may still infer what properties may affect anomalous diffusion by considering the descriptor subset selection of the LASSO regression. \\
\indent The selected descriptors for $\log{D_{P}^{(\alpha\beta)}}$ suggest that moderately polarizable and compact molecules diffuse relatively freely through PDMS, but those with certain functional groups (like amines) or locally charged regions may diffuse more slowly. For $\alpha$, more polar and electronically active molecules show more Fickian diffusion, while for $\beta$, more hydrophobic or polar molecules demonstrate more non-Gaussian diffusion while molecules with moderate polarity (or amphiphilicity), like those with hydroxyl or amino groups show Gaussian diffusion. While these results are highly exploratory and do not represent a predictive model, they suggest generally that molecule electrotopology may mediate anomalous diffusion in PDMS.

\subsection*{Integrating STF diffusion in mass transport modeling}
\begin{figure}
    \centering
    \includegraphics[scale = 0.725]{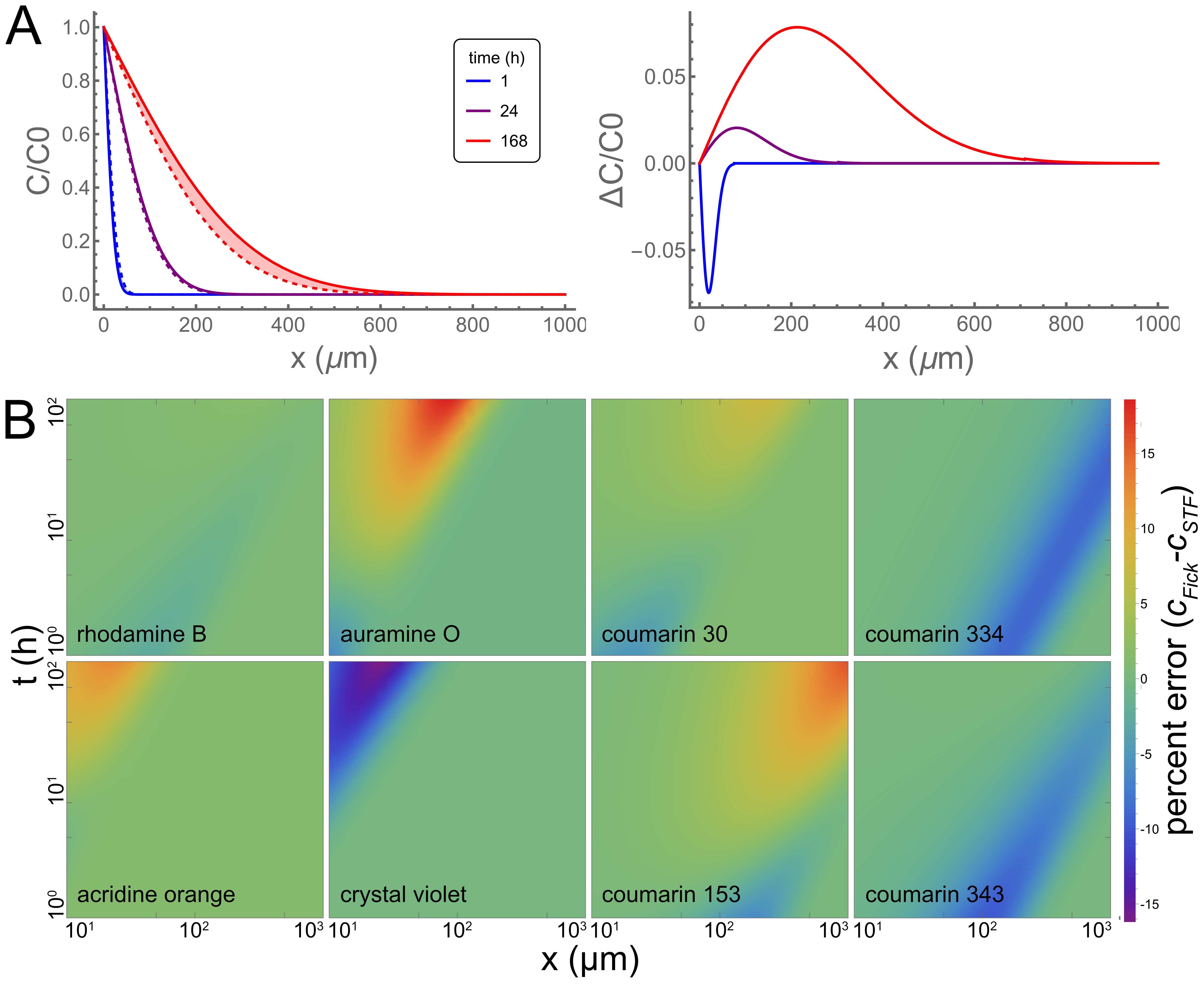}
    \caption{(A-left) Comparison of coumarin 30 diffusion in 1000 \textmu m-thick PDMS at 1 (blue), 24 (purple), and 168 (red) hours under the assumptions of Fickian (solid) and STF (dashed) diffusion. Note that coumarin 30 was found to have a Fickian diffusivity of 165.9 \textmu m\textsuperscript{2}/h. STF parameters of coumarin 30 diffusivity are found in Table \ref{STFFits}. (A-right) Calculated values of $\Delta c/c_{0} = (c_{Fick}-c_{STF})/c_{0}$ over 1 week.}
    \label{fig:FickvsSTF}
\end{figure}

 To extend this model to systems with anomalous STF diffusion, we first introduce a new diffusion equation for concentration $c_{P}$ in PDMS using the Riemann-Liouville fractional time derivative ($\mathcal{D}_{t}^{\beta}$), along with a standard Fickian diffusion equation for concentration in solution $c_{S}$:
\begin{align}
    \frac{\partial c_{P}}{\partial t} &= \frac{\alpha}{\beta}t^{\frac{\alpha}{\beta}-1}D_{P}^{\alpha\beta} \ _{RL}\mathcal{D}_{t^{\alpha/\beta}}^{1-\beta}\nabla^{2}c_{P} \\
    \frac{\partial c_{S}}{\partial t} &= D_{S}\nabla^{2}c_{P}
\end{align}
We must also modify the flux boundary conditions at the solution-PDMS interface ($\partial\Omega$) to account for STF diffusion and conserve mass:
\begin{align}
    _{RL}\mathcal{D}_{t}^{1-\alpha}D_{P}^{(\alpha\beta)}\mathbf{\hat{n}}\cdot\mathbf{\nabla}c_{P}\rvert_{\partial\Omega} &= +H(\kappa_{PS}c_{S}-c_{P}) \\
    D_{S}\mathbf{\hat{n}}\cdot\mathbf{\nabla}c_{S}\rvert_{\partial\Omega} &= -H(\kappa_{PS}c_{S}-c_{P})
\end{align}
For a full description of these operators, please see Part I of this work. Analytic methods are not generally available for this system, so one must implement numerical methods to develop approximate solutions. \\
\indent We may however consider simplified systems in which we may use the translational approach developed in Part I to develop solutions to STF diffuion. Consider a PDMS-based microfluidic device. Such devices have had mass transport comprehensively modeled under the assumption of Fickian diffusion \cite{Hermann2025}. In these devices, solutions of chemicals are pumped through small microchannels with velocity profiles given by laminar Poiseuille flow. If the aqueous diffusion rate in these channels is greater than the flow rate (that is, the Péclet number $\text{Pe}<1$), this system is diffusion-dominated and any cross-sectional slice of the channel has approximately uniform solution concentration. In a system with Fickian diffusion in PDMS, the concentration profile in PDMS for chemicals which have partitioned into PDMS at position $z$ along the channel $c_{P}(x,t)$ is scales with the complementary error function,
\begin{equation}
    c_{P}(x,t) = \kappa_{PS}c_{S}\text{erfc}\left(\frac{x}{\sqrt{4D_{P}t}}\right)
\end{equation}
where $\kappa_{PS}$ is the PDMS-solution partition coefficient and $c_{S} = c_{S}(z,t)$ is the concentration in solution. Using the translation approach developed in Part I of this work, in a system with STF diffusion, the concentration profile scales with the \textit{fractional} complementary error function (K),
\begin{equation}
    c_{P}(x,t) = \kappa_{PS}c_{S}K_{\beta/2}\left(\frac{x}{\sqrt{D_{P}^{(\alpha\beta)}t^{\alpha}}}\right)
\end{equation}
We will normalize these solutions ($\kappa_{PW}=1$, $c_{W}=1$) to consider the potential impact of cross-talk between channels in a device. Cross-talk occurs when chemical leeches from one channel and partitions out into solution in another channel in which the chemical should not be present. Comparison of simulated profiles for Fickian and STF diffusion are shown in Fig. \ref{fig:FickvsSTF}A for coumarin 30 at times from 1 to 168 h (1 week) for a 1000 \textmu m thick slab of PDMS. More generally, the percent error between a Fickian approximation and the STF model is shown in Fig. \ref{fig:FickvsSTF} for all eight dyes, revealing spatiotemporal domains in which the Fickian approximation may over- or under-fit the STF model. This error may be up to $\sim 20\%$ in magnitude. In general, to cover the same distance, the STF diffusive process will take some duration $t_{STF} = t^{1/\alpha}_{Fick}$ relative to the characteristic Fickian time $t_{Fick}$. Depending on this duration, and the value of $\alpha$, one may have shorter or longer than expected characteristic STF times. Depending on device geometry and the sensitivity of an in-device assay to chemical concentration, this may be relevant to the use of these devices and the modeling thereof.

\section*{Conclusion}
A selection of eight fluorescent dyes, spanning a wide range of physical parameters (mass, volume, polarity, and hydrophobicity) exhibit anomalous diffusion in bulk PDMS. This behavior is well-described by stretched-time fractional diffusion, an ensemble behavior corresponding to a stochastic, single-molecule process of generalized gray Brownian motion. In PDMS-based microfluidic devices, this behavior can lead to important physical consequences such as cross-talk between nearby channels. Given the relevance of such devices in biomedical sciences and engineering, characterizing anomalous diffusion through mass transport models with integrated STF diffusion will allow researchers to better model in-device chemical concentrations and accurately construct dose-response relationships for pharmacological or toxicological assays.

\section*{Data Availability Statement}
Data for this article, including the raw videos of dye diffusion and associated profiles s, are available at Open Science Framework at \url{https://osf.io/yh6wb/}. 

\section*{Conflicts of interest}
There are no conflicts of interest to declare.

\section*{Acknowledgments}
This publication was supported by U.S. Environmental Protection Agency (EPA) STAR Center
Grant \#84003101. Its contents are solely the responsibility of
the grantee and do not necessarily represent the official views of the U.S. EPA. Further, U.S. EPA does not endorse the purchase of any commercial products or services mentioned in
the publication.

\section*{Author Contributions}
Conceptualization, N.G.H., M.S.H.; methodology, N.G.H., D.A.M., M.S.H.; software, N.G.H.; validation, N.G.H.; formal analysis, N.G.H; investigation, N.G.H.; resources, N.G.G.; data curation, N.G.H.; writing---original draft preparation, N.G.H.; writing---review and editing, N.G.H., D.A.M., and M.S.H.; visualization, N.G.H.; supervision, D.A.M. and M.S.H.; project administration, M.S.H.; funding acquisition, D.A.M. and M.S.H. All authors have read and agreed to the published version of the manuscript.

\appendix
\section*{Appendix}
\section{Diffusion with error-function-like initial condition}
\begin{figure}[]
    \centering
    \includegraphics[scale = 0.9]{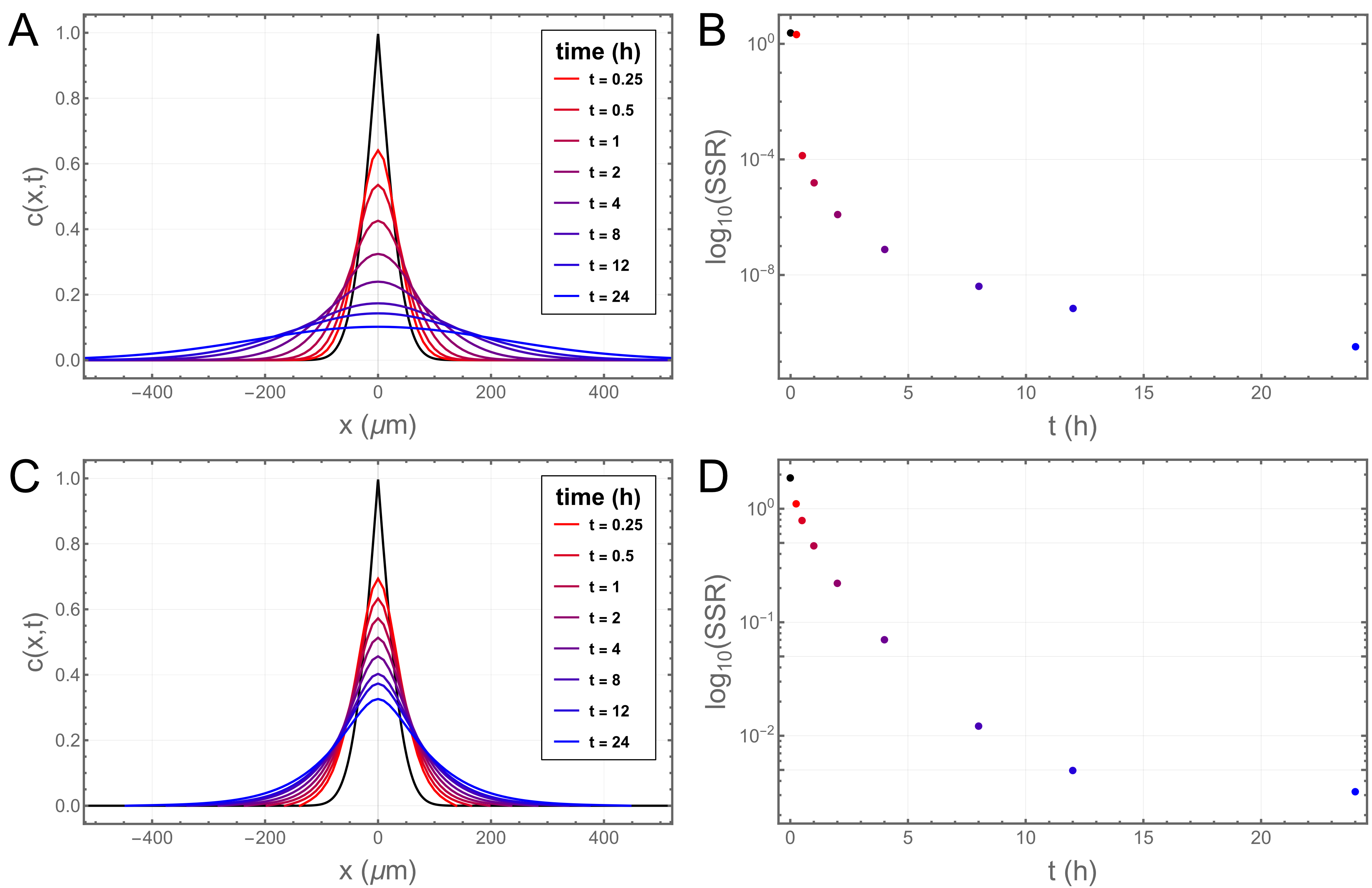}
    \caption{(A) The convolution of a Gaussian with a symmetrized error function initial condition (black) approaches a Gaussian solution over time. Here, $c_{0} = 1$ M, $D_{P} = 1000$ \textmu m\textsuperscript{2}/h and $\eta_{0} = 50$ \textmu m. (B) Summed-square residuals of a fit of Gaussian (Eqn. A.2) to numerical convolution decrease at each time. (C) The convolution of a M-Wright function with a symmetrized generalized error function ($K$) initial condition (black) approaches an M-Wright solution over time. Here, $c_{0} = 1$ M, $D_{P}^{(\alpha\beta)} = 1000$ \textmu m\textsuperscript{2}/h\textsuperscript{$\alpha$}, $\alpha=\beta=0.75 $ and $\eta_{0} = 50$ \textmu m. (D) Summed-square residuals of a fit of M-Wright function (Eqn. A.4) to numerical convolution decrease at each time.}
    \label{fig:CLT}
\end{figure}

If a chemical exhibits have Fickian diffusion in 1D over $\mathbb{R}$, with initial condition 
\begin{equation}
    c(x,0) = c_{0}\text{erfc}\left(\frac{x}{\eta_{0}}\right)
\end{equation}
the shape of the profile at time $t$ will approach a Gaussian shape; however it will not be exactly Gaussian given that Eqn. 16 lacks compact support. This result follows mathematically from the convolution of the initial condition with the appropriate Green's function:
\begin{equation}
    c(x,t) = \int_{-\infty}^{+\infty}\frac{c_{0}}{\sqrt{4\pi D_{P}t}}\exp{\left(-\frac{(x-x')^{2}}{4D_{P}t}\right)}\text{erfc}\left(\frac{x'}{\eta_{0}}\right)dx'
\end{equation}
The approximate concentration profile has shape
\begin{equation}
    c(x,t) = A\exp{\left(-\frac{x}{\sigma^{2}(t)}\right)}
\end{equation}
where $\sigma^{2}(t) = 4Dt + \sigma_{0}^{2}$. Here, $\sigma_{0}$ is a parameter corresponding to an idealized Gaussian initial condition. Numerical integration with physically realistic Fickian diffusivity, $D = 1000$ \textmu m\textsuperscript{2}/h, and an initial condition defined by $c_{0} = 1$ (a normalized profile) with $\eta_{0} = 50$ \textmu m, shows that we recover Gaussian behavior quickly. As shown in Fig. \ref{fig:CLT}A-B, the numerical convolution is increasingly well fit by a Gaussian approximation.. \\
\indent If a chemical instead exhibits anomalous STF diffusion and thus has an initial condition given by a fractional complementary error function
\begin{equation}
    c(x,0) = c_{0}\ K_{\beta/2}\left(\frac{|x|}{\eta_{0}}\right)
\end{equation}
then its evolution through STF diffusion follows the convolution with the corresponding Green's function involving the M-Wright function:
\begin{equation}
    c(x,t) = \int_{-\infty}^{+\infty}\frac{c_{0}}{\sqrt{4D_{P}^{(\alpha\beta)}t^{\alpha}}}M_{\beta/2}\left(\frac{|x-x'|}{\sqrt{D_{P}^{(\alpha\beta)}t^{\alpha}}}\right)K_{\beta/2}\left(\frac{|x'|}{\eta_{0}}\right)dx'
\end{equation}
In this case, the shape of the concentration profile at time $t$ will approach an M-Wright function with shape dependent on $\beta$,
\begin{equation}
    c(x,t) = A\ M_{\beta/2}{\left(\frac{|x|}{\sigma(t)}\right)}
\end{equation}
As shown in Fig. \ref{fig:CLT}C-D, this numerical convolution is increasingly well fit by an M-Wright function (Eqn. A.6). 
This characteristic M-Wright shape is a consequence of the Generalized Central Limit Theorem, which is satisfied by distributions of the M-Wright type, as these distributions are Lévy stable distributions \cite{Mainardi2010aF,Pagnini2012,Amir2020}.

\bibliographystyle{ieeetr}
\bibliography{citations}

\end{document}


\noindent\LARGE{\textbf{Anomalous diffusion for mass transport phenomena II: Subdiffusion in polydimethylsiloxane (PDMS) \\ \textit{Supplement}}}\\

 \noindent\large{Nathaniel G. Hermann,\textit{$^{a}$}  Dmitry A. Markov,\textit{$^{b}$} and M. Shane Hutson\textit{$^{a}$}}$^{\ast}$ \\
\textit{$^{a}$~Department of Physics and Astronomy, Vanderbilt University }\\
\textit{$^{b}$~Department of Biomedical Engineering, Vanderbilt University }
\\

\renewcommand{\thetable}{S\arabic{table}}

\begin{table}[h]
  \centering
  \caption{Molecular descriptors linearly correlated with $\log{D_{P}^{(\alpha\beta)}}$, showing LASSO regression coefficients $\beta_{j}$ for descriptor $j$ and $R^2$ values from univariate linear fits.}
  \vspace*{1mm}
  \label{tbl:example2}
  \begin{tabular}{lcc}
    \hline
    Descriptor & $\beta_{j}$ & $R^2$ \\
    \hline
    MaxEStateIndex & 0.481 & 0.859 \\
    MaxAbsEStateIndex & 0.061 & 0.859 \\
    FpDensityMorgan1 & 0.016 & 0.836 \\
    SMR\_VSA5 & 0.144 & 0.889 \\
    SMR\_VSA6 & -0.170 & 0.815 \\
    VSA\_EState9 & -0.095 & 0.903 \\
    fr\_NH0 & -0.578 & 0.761 \\
    \hline
  \end{tabular}
\end{table}

\begin{table}[h]
  \centering
  \caption{Molecular descriptors linearly correlated with STF diffusion order $\alpha$, showing LASSO regression coefficients $\beta_{j}$ for descriptor $j$ and $R^2$ values from univariate linear fits.}
  \vspace*{1mm}
  \label{tbl:example2}
  \begin{tabular}{lcc}
    \hline
    Descriptor & $\beta_{j}$ & $R^2$ \\
    \hline
    PEOE\_VSA7 & 0.003 & 0.592 \\
    VSA\_EState7 & 0.042 & 0.801 \\
    \hline
  \end{tabular}
\end{table}

\begin{table}[h]
  \centering
  \caption{Molecular descriptors linearly correlated with STF diffusion order $\beta$, showing LASSO regression coefficients $\beta_{j}$ for descriptor $j$ and $R^2$ values from univariate linear fits.}
  \vspace*{1mm}
  \label{tbl:example2}
  \begin{tabular}{lcc}
    \hline
    Descriptor & $\beta_{j}$ & $R^2$ \\
    \hline
    MinAbsEStateIndex & 0.015 & 0.446 \\
    BCUT2D\_LOGPHI & -0.003 & 0.461 \\
    PEOE\_VSA11 & -0.021 & 0.565 \\
    PEOE\_VSA12 & -0.021 & 0.367 \\
    EState\_VSA7 & 0.050 & 0.632 \\
    VSA\_EState7 & -0.11 & 0.473 \\
    NHOHCount & 0.019 & 0.379 \\
    fr\_NH2 & 0.008 & 0.409 \\
    \hline
  \end{tabular}
\end{table}

\newpage

\renewcommand{\arraystretch}{1.3}
\begin{table}[h]
  \centering
  \caption{Select molecular descriptors with physical interpretation and citations.}
  \vspace*{1mm}
  \label{tbl:example2}
  \begin{tabularx}{\textwidth}{lXc}
    \hline
    Descriptor & Interpretation & Citation \\
    \hline
    MaxEStateIndex & Maximum electronic state index (ESI) for single atom. Higher values correspond to higher electronic activity. & \cite{Kier1990} \\
    MaxAbsEStateIndex & Maximum absolute value of ESI for single atom. By focusing on magnitude, it identifies the most extreme local environment. & \cite{Kier1990} \\
    FpDensityMorgan1 & Density of Morgan fingerprint bits considering atoms and their nearest neighbors (radius = 1). Higher values suggest molecular complexity. & \cite{Morgan1965} \\
    SMR\_VSA5 & Sum of van der Walls surface area (VSA) for atoms with SMR contributions in bin 5 (SMR $\in [2.45,2.75]$). Intended to capture polarizability. & \cite{Labute2000} \\
    SMR\_VSA6 & Sum of VSA for atoms with SMR contributions in bin 6 (SMR $\in [2.75,3.05]$). Intended to capture polarizability. & \cite{Labute2000} \\
    VSA\_EState9 & Sum of VSA values for atoms with ESI in bin 9 (ESI $\in[4.69,9.17]$). Intended to capture the surface area associated with particular electronic characteristics.  & \cite{Kier1990,Labute2000} \\
    fr\_NH0 & Number of amines not part of an aromatic ring. & -- \\
    PEOE\_VSA7 & Sum of VSA for atoms with partial equalization of orbital electronegativities (PEOE) contributions in bin 7 (PEOE $\in [-0.05,0]$). Intended to capture direct electrostatic interactions and reflects partial charge distribution. & \cite{Labute2000} \\
    VSA\_EState7 & Sum of VSA values for atoms with ESI in bin 7 (ESI $\in[1.807,2.05]$). Intended to capture the surface area associated with particular electronic characteristics. & \cite{Kier1990,Labute2000} \\
    MinAbsEStateIndex & Minimul absolute value of ESI for single atom. By focusing on magnitude, it identifies the least polar local environment. & \cite{Kier1990} \\
    BCUT2D\_LOGPHI & A Burden eigenvalue (BCUT) based on weighted atomic Crippen logP contributions. Higher values suggest higher propensity for affinity with local hydrophobic environments. & \cite{Burden1989,Stanton1999} \\
    PEOE\_VSA11 & Sum of VSA for atoms with PEOE contributions in bin 11 (PEOE $\in [0.15,0.2]$). Intended to capture direct electrostatic interactions and reflects partial charge distribution. & \cite{Labute2000} \\
    PEOE\_VSA12 & Sum of VSA for atoms with PEOE contributions in bin 12 (PEOE $\in [0.2,0.25]$). Intended to capture direct electrostatic interactions and reflects partial charge distribution. &\cite{Labute2000} \\
    EState\_VSA7 & Sum of ESI values for atoms with VSA in bin 7 (VSA $\in[6.07,6.45]$). Intended to capture the surface area associated with particular electronic and topological characteristics. & \cite{Kier1990,Labute2000} \\
    NHOHCount & Number of N-O or O-H bounds in the molecule. May correspond to hydrogen bond donors. & -- \\
    fr\_NH2 & Number of primary amino groups in the molecule. & -- \\
    \hline
  \end{tabularx}
\end{table}

\newpage

\renewcommand{\thefigure}{S\arabic{figure}}

\begin{landscape}
\begin{figure}
    \centering
    \includegraphics[scale = 1.1]{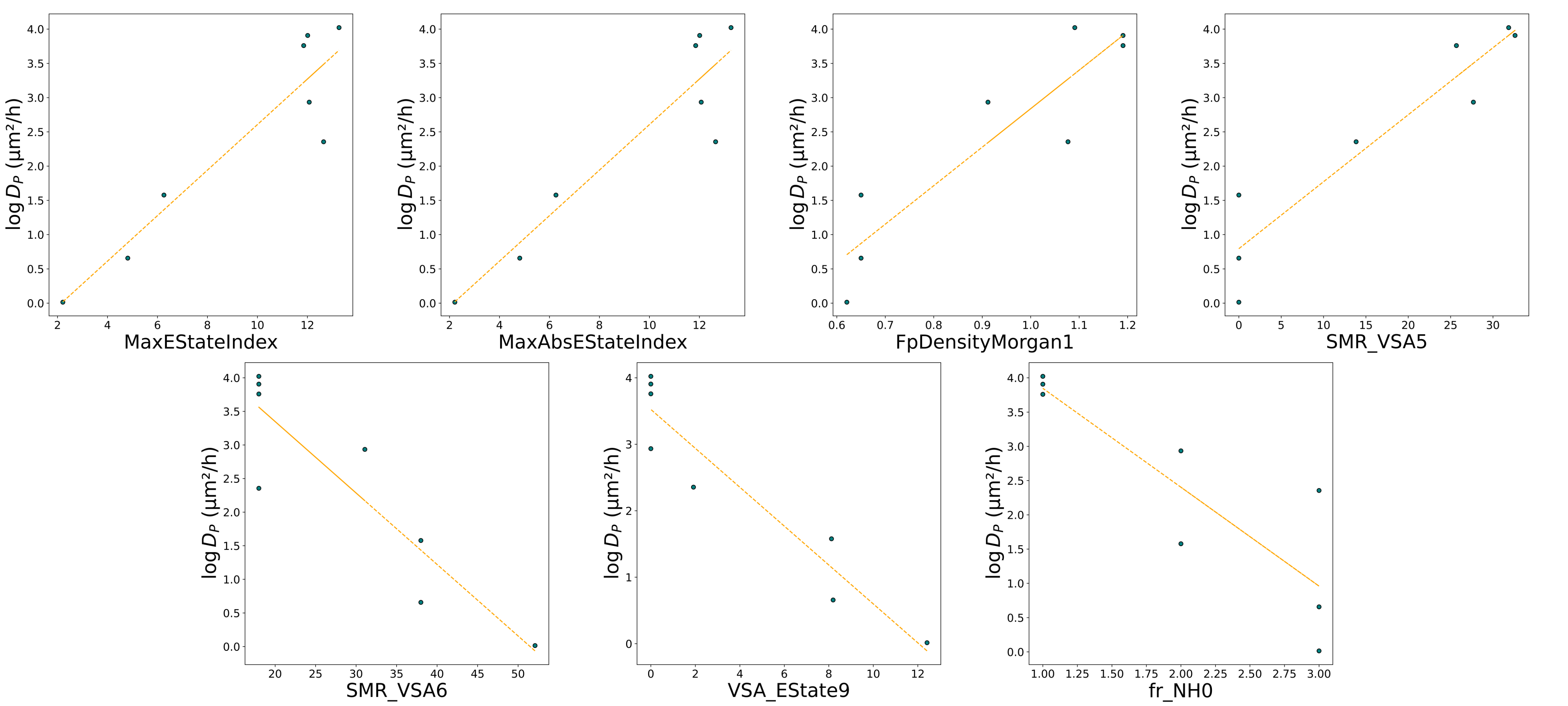}
    \caption{}
    \label{fig:test}
\end{figure}
\end{landscape}

\newpage

\begin{landscape}
\begin{figure}
    \centering
    \includegraphics[scale = 1.5]{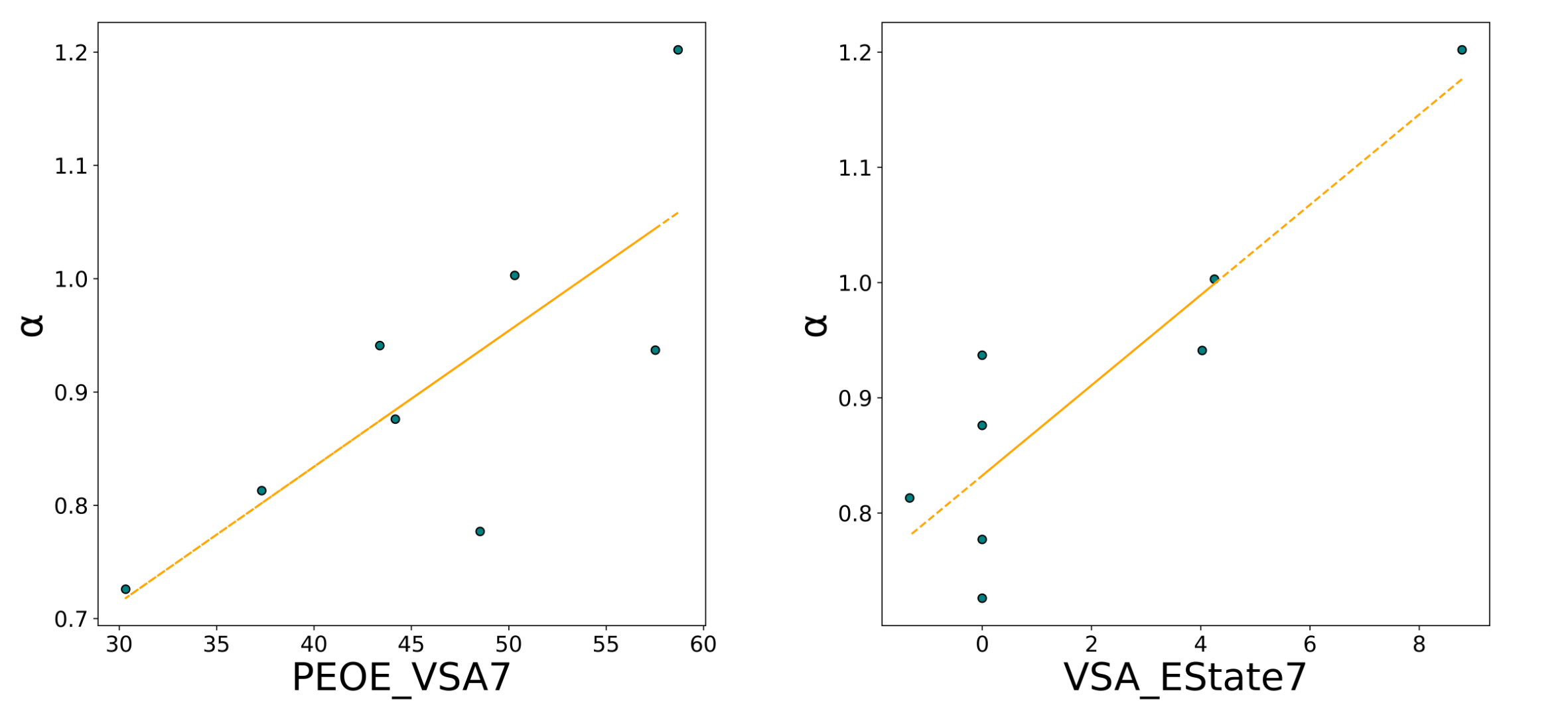}
    \caption{}
    \label{fig:test}
\end{figure}
\end{landscape}

\newpage

\begin{landscape}
\begin{figure}
    \centering
    \includegraphics[scale = 1.1]{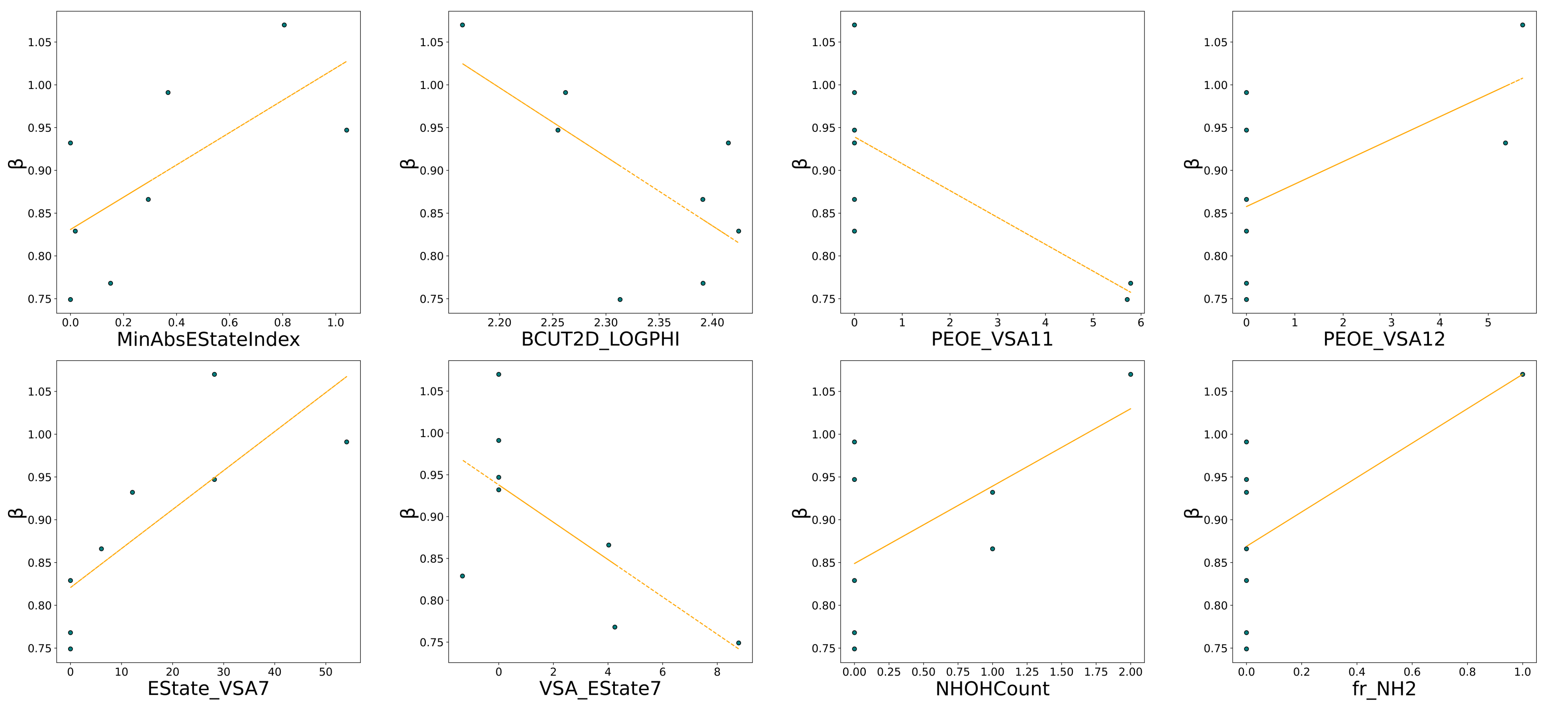}
    \caption{}
    \label{fig:test}
\end{figure}
\end{landscape}

\clearpage

\noindent\LARGE{\textbf{Supplemental Code}} \\

\renewcommand{\thelstlisting}{S\arabic{lstlisting}}

\small{
\begin{lstlisting}[language=Python, caption={Function to calculate molecular descriptors using RDKit.},frame=shadowbox]
import numpy as np
import pandas as pd
from rdkit import Chem
from rdkit.ML.Descriptors import MoleculeDescriptors
from rdkit.Chem import Descriptors
def desc_generator(chem_names,smiles_list):
    # Molecules from SMILES
    mols = [Chem.MolFromSmiles(smi) for smi in smiles_list]
    # Get all RDKit descriptors
    desc_names = [desc[0] for desc in Descriptors._descList]
    calc = MoleculeDescriptor.MolecularDescriptorCalculator(desc_names)
    # Descriptors from molecules
    desc_data = [calc.CalcDescriptors(mol) for mol in mols]
    desc_df = pd.DataFrame(desc_data,columns=desc_names,index=chem_names)
    return desc_df
\end{lstlisting}
}

\small{
\begin{lstlisting}[language=Python, caption={Function to find relevent descriptors for STF diffusion parameters using LASSO regression.},frame=shadowbox]
import numpy as np
import pandas as pd
from sklearn.linear_model import LassoCV, LinearRegression
from sklearn.preprocessing import StandardScaler
from sklearn.pipeline import make_pipeline
from sklearn.metrics import r2_score
from sklearn.model_selection import LeaveOneOut, cross_val_predict
def important_desc(exp_df,desc_df,param):
    # Loading values of RDKit descriptors and target parameter
    X = desc_df.values
    y = exp_df[param].values #e.g. param='alpha'
    # Setting up pipeline with standard scaling and LOO validation
    model = make_pipeline(StandardScaler(),
        LassoCV(cv=LeaveOneOut(),random_state=1,max_iter=10000)
        )
    model.fit(X,y)
    # Calcuate model r-squared
    r_squared = model.score(X, y)
    print(f"R-squared: {r_squared:.3f}")
    # Calculate cross-validated model r-squared
    y_cv_pred = cross_val_predict(model, X, y, cv=LeaveOneOut())
    r2_cv = r2_score(y, y_cv_pred)
    print(f"Cross-validated R-squared: {r2_cv:.3f}")
    # Query for important coefficients
    coef = model.named_steps['lassocv'].coef_
    important_desc = [name for name, c in zip(desc_df.columns,coef) if c!=0]
    coef = coef[coef!=0]
    # Get univariate R2 scores
    r2_scores = []
    for desc in important_desc:
        X_single = desc_df[[desc]].values
        model = LinearRegression().fit(X_single,y)
        r2_scores.append(r2_score(y,model.predict(X_single)))
    # Summary dataframe
    summary_df = pd.DataFrame({
        'Descriptor': important_desc,
        'Lasso Beta': coef,
        'R2': r2_scores
    })
    return summary_df
\end{lstlisting}
}

\bibliographystyle{unsrt}
\bibliography{citations}